\documentclass[graybox]{svmult}
\usepackage{geometry}

\usepackage{type1cm}        
\usepackage{makeidx}         
\usepackage{graphicx}        
\usepackage{multicol}        
\usepackage[bottom]{footmisc}

\usepackage{newtxtext}       %
\usepackage[varvw]{newtxmath}       

\makeindex             

\usepackage{amsmath,amsfonts,amsthm}
\usepackage{graphicx,xcolor} 
\usepackage{comment,bm}
\usepackage[english]{babel}

\usepackage{algorithm}
\usepackage{textcomp}
\usepackage{algpseudocode}
\usepackage{bbm}
\usepackage[shortlabels]{enumitem}
\usepackage{wrapfig}
\usepackage{tabularx,xcolor}
\usepackage{multirow}

\newcommand{\edit}[1]{\textcolor{black}{#1}}

\begin{document}

\title*{\edit{A non-parametric optimal design algorithm for population pharmacokinetics}}

\author{Markus Hovd \orcidID{0000-0002-6077-0934}
and\\ Alona Kryshchenko \orcidID{0000-0002-5049-0293}
and\\ Michael N. Neely \orcidID{0000-0002-1675-8276}
and\\ Julian Otalvaro \orcidID{0000-0001-5202-1645}
and\\ Alan Schumitzky 
and\\ Walter M. Yamada \orcidID{0000-0003-3512-9202}}
\authorrunning{Markus Hovd, Alona Kryshchenko, Michael N. Neely, Julian Otalvaro, Alan Schumitzky, Walter M. Yamada} 

\institute{Markus Hovd \at Oslo University Hospital, University of Oslo \email{markus.hovd@farmasi.uio.no}
\and Alona Kryshchenko \at California State University Channel Islands \email{alona.kryshchenko@csuci.edu}
\and Michael N. Neely \at Children's Hospital Los Angeles, University of Southern California  \email{mneely@chla.usc.edu }
\and Julian Otalvaro \at Children's Hospital Los Angeles \email{jotalvaro@chla.usc.edu} 
\and Alan Schumitzky \at University of Southern California \email{schumitzky@gmail.com}
\and Walter M. Yamada \at Children's Hospital Los Angeles \email{wyamada@chla.usc.edu }}

\maketitle

\abstract{ This paper introduces a non-parametric estimation algorithm designed to effectively estimate the joint distribution of model parameters with application to population pharmacokinetics. Our research group has previously developed the non-parametric adaptive grid (NPAG) algorithm, which while accurate, explores parameter space using an ad-hoc method to suggest new support points. In contrast, the non-parametric optimal design (NPOD) algorithm uses a gradient approach to \edit{suggest} new support points\edit{, which reduces the amount of time spent evaluating non-relevant points and by this the overall number of cycles required to reach convergence}. \edit{In this paper, we demonstrate that the NPOD algorithm achieves similar solutions to NPAG across two datasets, while being significantly more efficient in both the number of cycles required and overall runtime. Given the importance of developing robust and efficient algorithms for determining drug doses quickly in pharmacokinetics, the NPOD algorithm represents a valuable advancement in non-parametric modeling. Further analysis is needed to determine which algorithm performs better under specific conditions.}}

\section{Introduction}
Pharmacokinetic modeling and simulation have become a cornerstone in both drug development and therapeutic drug monitoring. The ability to integrate pre-clinical and clinical data, along with covariates, allows for accurate inference of both drug exposure (pharmacokinetics, PK) and response (pharmacodynamics, PD). These statistics are integral to drug therapy optimization at both the individual and population levels. Two different statistical approaches are common: parametric and non-parametric \cite{Goutelle2022}. While parametric approaches assume that the probability distribution of model parameter values follows predefined distributions such as the normal and log-normal \cite{DArgenio09, shah2019novel, Ishihara2020, Soraluce2020, Allard2020}, non-parametric approaches are free of this assumption. Rather, the joint parameter value probability distribution consists of discrete support points, each point comprising a vector of values for every parameter and an associated probability based on the likelihood of those parameter values. If desired, the shape of the distribution can be inferred by calculations on the optimized support points and their corresponding probabilities, e.g. covariance, mean or median, or an operation on the points for example kernel density estimation. Non-parametric approaches allow for more accurate a priori detection of sub-populations and outliers \cite{Neely2012, Goutelle2022a}.

 The non-parametric adaptive grid (NPAG) algorithm is a well-established non-parametric estimation method widely used in pharmacokinetics and pharmacodynamics (PK/PD) \cite{Goutelle2022a}. NPAG is a \edit{``}throw and catch\edit{''} algorithm. It begins with a viable solution (that is, the likelihood of the current set of support points is greater than 0), and assuming a better solution can be found on the Euclidean grid surrounding each support point of that viable solution, it casts out new, potentially better support points along each dimension of the grid. Successive cycles will find the local optimum around the viable solution. Confidence is gained as the grid is \edit{``}adaptive\edit{''} in both discretization length and position in space. However, due to the nature of the adaptive grid, NPAG is computationally expensive, and therefore slow to converge.

With increasingly complex PK-PD models, large in the number of parameters, subjects, or both, algorithm speed becomes critical. This has motivated the development of the current non-parametric estimation technique that can maintain the accuracy of NPAG while significantly improving time to convergence. Addressing the convergence speed issue in non-parametric estimation is crucial for streamlining the PK/PD modeling and analysis workflow, enabling faster and more cost-effective drug development processes. The proposed algorithm in this paper tackles this challenge by introducing innovative computational methods and optimization strategies to enhance the efficiency of non-parametric parameter estimation.

\section{Methods}
\subsection{Design of the Non-parametric Optimal Design Estimation Algorithm}

Pharmacokinetic observations can be statistically described using a mixing distribution model, where the probability of random variable arguments (the PK population model) in the PK compartmental model is governed by a mixing distribution.

The task of estimating this mixing distribution from a set of PK observations can be defined as follows. Let ${Y}_1, ..., {Y}_N$ represent a sequence of independent but not necessarily identically distributed random vectors, constructed from one or more observations from each of $N$ subjects in the population. Additionally, let ${\theta}_1,...,{\theta}_N$ denote a sequence of independent and identically distributed random vectors representing unknown parameter values for $N$ subjects. These ${\theta}$ values belong to a compact subset $\Theta$ of Euclidean space with a common but unknown distribution $F$, representing the parameter space of the population model.

The objective is to maximize the likelihood function $L(F)$ with respect to all probability distributions $F$ on $\Theta$. Each ${\theta}_i$ is not observed, but it is assumed that the conditional densities $p({Y}_i \vert {\theta}_i)$ are known for $i = 1,..., N$. The mixing distribution of ${Y}_i$ with respect to $F$ is then given by $p({Y}_i \vert F ) = \int p({Y}_i \vert {\theta}_i ) dF({\theta}_i)$. 

Let $F^{ML}$ be the distribution that maximizes $L(F)$. It serves as a consistent estimator of the true mixing distribution. Because of independence of the $\{Y_{i}\}$, the likelihood function can be written as 

\begin{equation} \label{eq:1}
L(F)=p(Y_{1},...,Y_{N}|F)=\prod_{i=1}^{N}\int p(Y_{i}|\theta_{i})dF(\theta_{i})
\end{equation} 

It is important to note that $L(F)$ is a convex function of $F$. Further, it is shown in \cite{Lindsay1983}, under simple hypotheses, that the global maximizer $F^{ML}$ of $L(F)$ is a discrete distribution with at most $N$ support points, where $N$ is the number of subjects in the population and a support point is a vector of model parameter values with nonzero probability.

The problem of finding $F^{ML}$ has been addressed in \cite{Yamada2021} by the NPAG algorithm that uses the primal-dual interior point method to find optimal weights and an Adaptive Grid algorithm to find optimal support points. It was also addressed in Lesperance and Kalbfleisch \cite{Lesperance1992} by the combination of the Semi-Infinite Programming algorithm to find optimal weights and the Improved Supervised Descent Method (ISDM) algorithm using the D-function to find optimal support points. And in \cite{Wang2015} by the combination of Quadratic Programming algorithm to find optimal weights and ISDM algorithm to find optimal support points.

The algorithm described here is an alternative to the NPAG algorithm and is different from it in the step of finding optimal support points. It utilizes the primal-dual interior-point method for convex programming to find optimal weights of the $F^{ML}$ and introduces the optimization of the directional derivative of the likelihood function to address the search for optimal support points of the $F^{ML}$. This algorithm was proposed by Dr. Robert Leary in the PAGE conference poster \cite{leary2007evolutionary}.

The design of the NPOD algorithm is summarized in the steps below.
\subsubsection{Design principles and theoretical foundation}
Traditionally, non-parametric maximum likelihood methods rely on iterative approaches such as the expectation-maximization algorithm, which entails optimizing conditional expectation. However, this process can be quite time-intensive, particularly for problems with high dimensions. To address this, we've developed an enhanced iterative non-parametric optimal design (NPOD) algorithm that streamlines certain optimization stages using directional derivatives, significantly boosting its speed compared to the original version detailed by \cite{Neely2012, Yamada2021}.

\subsubsection{Algorithm Implementation}
\paragraph{\textit{Initialization:}}
The first step of any non-parametric algorithm is the initialization of the n-dimensional parameter space. Importantly, the parameter space must be bounded, as an infinitely large parameter space is both computationally and physiologically impossible. In most cases, the sample space represents an uninformed prior. However, it may also be initialized with the joint distribution obtained from previous searches, or other algorithms. In the present implementation of NPOD we are using a modified version of the Sobol pseudo-random sequence generator based on the work by Burley et al \cite{burley_practical_2020} with an improved hash by Kuo et al \cite{joe_remark_2003, joe_constructing_2008} 

\paragraph{\textit{Likelihood Calculation:}}
The next step is the calculation of the likelihood or the objective function. This is the most computationally expensive step in the algorithm, as it is calculated by solving the differential equation representing the pharmacokinetic model for each subject for each point in the initial grid.

\paragraph{\textit{Optimization:}}
Following the calculation of the likelihood, the weight of each support point is recalculated in order to maximize the sum of the likelihood function across all subjects. This is achieved through the use of a primal-dual interior point algorithm \cite{Yamada2021}.

\paragraph{\textit{Rank revealing function:}}
Another important property of the joint parameter distribution in the non-parametric approach is that the maximum number of support points can at most be equal to the number of subjects. Non-optimal solutions can have more support points than the number of subjects. In this step, we use QR decomposition of the $\Psi = {P(Y_i|\theta_k)}_{N \times K}$ matrix and remove all the support points that are not in the orthonormal basis for the column space of $\Psi$ matrix. We do this at each cycle to guarantee optimizations never expand uncontrollably.

\paragraph{\textit{Support Point Adjustment:}}
It is at this step that the NPOD and NPAG algorithms diverge; while NPAG employs an adaptive grid to suggest new support points in the search space, NPOD employs a directional derivative of the log-likelihood function using Nelder-Mead algorithm \cite{cacf87ec-1180-3742-953b-557a4a81081b}. 

The directional derivatives of the log-likelihood of $F$ in the direction of the atomic density function centered at each support point is denoted as $D_{\delta_{\xi}}\ell(F)$. The idea originates in a text by Fedorov \cite{fedorov:toe72}, which covers $D$-optimal design theory. Another connection to Fedorov's $D$-optimal design theory and maximum likelihood estimators is provided by Mallet \cite{mallet:mlemrcrm86}. That paper provides an alternative to Lindsay's approach. In fact, according to Schumitzky \cite{schumitzky:nemaepd91} Lindsay and Mallet worked jointly to develop the theory that reduced the space of distributions to the space of only discrete distributions with $K$ support points, denoted $\mathcal{F}_K$ (where $K$ is no more than the number of subjects $N$). 

Let $F$ be any distribution on $\Omega$, the space of parameters for $\xi$. Then define the directional derivative D-Function as 
\begin{align}
    \label{eq:arab4.3.2} D(\xi, F) = \left( \sum \limits_{i=1}^N \frac{P(Y_i \mid \xi)}{P(Y_i \mid F)} \right) - N
\end{align}
where $\xi$ is a parameter and $N$ is the population size. Lindsay \cite{lindsay:gml83} showed that $F^* = F^{ML}$ if and only if 
\begin{align}
    \label{eq:arab4.3.3} \max \limits_{\xi \in \Omega} D(\xi, F^*) = 0.
\end{align}
Additionally, in the same paper, Lindsay showed when 
\begin{align}
    \label{eq:arab4.3.4} \max \limits_{\xi \in \Omega} D(\xi, F^*) \neq 0,
\end{align}
it is still true that 
\begin{align}
    \label{eq:arab4.3.5} L(F^{ML}) - L(F^*) \leq \max \limits_{\xi \in \Omega} D(\xi, F^*)
\end{align}
for $F^*, F^{ML} \in \mathcal{F}_K$.

In NPOD the updated set of support points is found as follows: for $k=1,...,K$ where $K$ is the current grid size and $F^{(n)}$ is the current distribution:

\begin{equation}
    \theta^{(n+1)}_k = argmax^{t}_{\xi \in \Omega}(D(\xi, F^{(n)})), \\
    \label{nes_sup}
\end{equation}
\begin{equation}
    D(\xi, F^{(n)}) = \left( \sum \limits_{i=1}^N \frac{P(Y_i \mid \xi)}{P(Y_i \mid F^{(n)})} \right) - N,\\
\end{equation}

\begin{equation}   
    P(Y_i\mid F^{(n)}) =
    \sum_{l=1}^K(w_l^{(n)} P(Y_i \mid \theta_l^{(n)}))
\end{equation}

where $argmax^{t}$ only takes $t$ steps in the Nelder-Mead optimization process \cite{cacf87ec-1180-3742-953b-557a4a81081b}. This adjustment plays a pivotal role in enhancing the efficiency of NPOD compared to NPAG, particularly in achieving convergence to local maxima.

The parameter $t$ is regarded as one of the hyperparameters that can be fine-tuned to optimize the performance of the algorithm. Typically, we set $t$ to be less than 5, based on empirical observations and computational experiments. This choice balances the trade-off between computational cost and optimization effectiveness.

By limiting the number of steps in the Nelder-Mead optimization, we can focus the algorithm's search on promising regions of the parameter space while avoiding excessive computational overhead. This targeted approach enables NPOD to converge more efficiently towards local maxima, making it a valuable tool for solving optimization problems in diverse domains.  

Once $\theta^{(n+1)}_k$ is determined through the optimization process, a validation step ensues to ensure its integrity within the algorithm. Specifically, it undergoes scrutiny to confirm two critical aspects: firstly, that it constitutes a distinct point, and secondly, that it remains within the predefined boundary conditions.

This validation mechanism serves as a safeguard against redundancy and boundary violations, both of which could potentially compromise the accuracy and reliability of the optimization process. By confirming the uniqueness and adherence to boundary constraints of $\theta^{(n+1)}_k$, the algorithm maintains the integrity of its parameter space exploration, facilitating robust and effective optimization outcomes.

\paragraph{\textit{Convergence:}}
The previous steps, excluding initialization, are iteratively repeated until no further improvement can be found, indicating convergence to an optimal solution. Improvement is evaluated by change in the likelihood, for which we consider a change less than $10^{-4}$ to indicate convergence. 

\subsubsection{Computational considerations and optimizations}
\paragraph{Initial search space:}

NPOD is initialized with a sufficiently compact set of support points within the search space. We report results for varying density Sobol sequence initializations in the Results section.

\paragraph{Hyperparameters:}
The NPOD algorithm, relying on the D-optimization function, is tuned by the number of iterations of the Nelder-Mead algorithm $t$. In our examples, a value of 5 was used for $t$, empirically chosen based on experience.

\subsection{Software implementation}
Recently, significant efforts has been placed in creating a new framework for pharmacometric algorithm development. While the original NPAG algorithm was written in Fortran, both the NPAG and NPOD algorithm has been rewritten in Rust, a memory-safe and computationally efficient programming language. While the framework itself will be presented in a future work, both algorithms are available to use in the development branch of the Pmetrics code repository \cite{PmetricsGitHub}. All computations were performed on a MacBook Pro (Apple) equipped with an M3 Max processor with 128GB of RAM.

\subsection{Comparative analysis of NPOD with NPAG}
The natural choice of a comparative algorithm for non-parametric pharmacokinetic modeling is NPAG. To compare the algorithms, we use two datasets, one synthetic in which the real parameter distribution is known and another, using real pharmacokinetic data from subjects in which the underlying distribution is not known. We will refer to these as datasets A and B, respectively.

The model used to fit dataset A is shown in Equation \ref{eq:bmke}, where $A$ is the amount of drug in the central compartment, $K_e$ is the elimination rate constant with a bimodal distribution, and $V_d$ is the apparent volume of distribution with unimodal distribution. The model includes an intravenous infusion, modeled as $R_{inf}$, the rate of infusion.

\begin{equation}
    \frac{dA}{dt} = -K_{e} \cdot A + R_{inf}, \quad C = \frac{A}{V_d}+\epsilon
    \label{eq:bmke}
\end{equation}

Dataset A consisted of simulated data with known parameter distribution, and with known measurement noise in the observations $\epsilon \sim \mathcal{N}(0, 0.05*C)$, previously discussed by Neely et al \cite{Neely2012}. It includes a total of 51 simulated subjects, all of whom received an intravenous infusion of 500 units over a duration of 30 minutes. Each subject was sampled 10 times over 24 hours, at 0.5, 1, 2, 3, 4, 6, 8, 12, 18, and 24 hours from the start of the infusion.

The model used to fit dataset B is shown in Equation \ref{eq:teqlag}, where $A_1$ represents the absorptive compartment and $A_2$ represents the amount of drug in the central compartment, from which $K_e$ is the elimination rate constant and $V_d$ is the apparent volume of distribution. The model includes an individual lag-term on the input dose $D$, modeled as a delayed unit Dirac delta function $\delta$.

\begin{equation}
    \frac{dA_1}{dt} = -K_a \cdot A_1 + D*\delta(t-t_{lag}) , \quad
\frac{dA_2}{dt} = K_a \cdot A_1 -K_{e} \cdot A_2 , \quad C = \frac{A_2}{V_d}+ \epsilon
    \label{eq:teqlag}
\end{equation}

Dataset B was originally provided as one of the example datasets available in the Pmetrics package for R \cite{Neely2012}. It includes data from 20 patients, all of whom received 600 units six times every 24 hours. A total of 139 samples were obtained across all subjects, all following the second-to-last dose. 

Any observation has an associated uncertainty, which must be accounted for during parameter estimation. We model uncertanty as $\epsilon$, which is normally distributed with mean zero and standard deviation $\omega$ defined by Eq. (\ref{eq:lambda}) or (\ref{eq:gamma}). First, an error polynomial model is used to estimate the uncertainty ($\sigma$) in each measurement ($y$). This is given in Equation \ref{eq:errorpoly}.

\begin{equation}
    \sigma = C_0 + C_1 \cdot y + C_2 \cdot y^2 + C_3 \cdot y^3
    \label{eq:errorpoly}
\end{equation}

Additional noise is modelled through either an additive ($\lambda$, Equation \ref{eq:lambda}) or proportional ($\gamma$, Equation \ref{eq:gamma}) error model. Each observation is then weighted by the reciprocal of the squared uncertainty, i.e. $1/\omega^2$.

\begin{equation}
    \omega = \sqrt{\sigma^2 + \lambda^2}
    \label{eq:lambda}
\end{equation}

\begin{equation}
    \omega = \sigma \cdot \gamma
    \label{eq:gamma}
\end{equation}

For the simulated dataset A, an additive error model was used with an initial value of $\lambda = 0$, and a flat uncertainty of 5\%, i.e. $C_1 = 0.05$, and $C_0 = C_2 = C_3 = 0$. For the real-world dataset B, a proportional error model was used with an initial value of $\gamma = 5$, and $C_1 = 0.02$, $C_2 = 0.05$, $C_3 = -0.002$ and $C_4 = 0$.

Both algorithms were compared on each dataset with various densities of the initial parameter search space, with otherwise equal conditions. Multiples of 51 was used for the number of initial support points, i.e. $K = 51 \cdot 2^{x}$, where $x$ ranged from 0 to 11, producing initial densities ranging from 51 to 104 448.

\section{Results}

 For dataset A, the location of support points at convergence for NPAG and NPOD, with an initial count of 104 448 support points, are illustrated in Figure \ref{fig:theta}. The weighted means for $K_e$ (NPAG = 0.187, NPOD = 0.187) and $V_d$ (NPAG = 103.7, NPOD = 103.8) between the two algorithms were almost identical. Furthermore, for all the different initial grid densities evaluated, ranging from 51 to 104 448, NPOD was able to achieve convergence at a much faster rate compared to NPAG, requiring almost one twentieth the number of cycles for a high number of initial points (Table \ref{tab:npod_npag}). However, the difference in overall computation time is negligible.

\begin{table}[H]
\centering
\begin{tabular}{r|cccc|cccc}
\hline
& \multicolumn{4}{c|}{NPAG} & \multicolumn{4}{c}{NPOD} \\
\cline{2-9}
\textnumero & Cycles & $LL$ & Support points & Time & Cycles & $LL$ & Support points & Time \\
\hline
51 & 196 & -646.45 & 48 & 3.87s & 21 & -646.85 & 47 & 1.56s \\
102 & 164 & -646.38 & 49 & 3.40s & 15 & -646.85 & 48 &  1.47s \\
204 & 132 & -646.52 & 48 & 3.00s & 13 & -646.80 & 47 & 1.47s \\
408 & 136 & -646.39 & 48 & 3.27s & 11 & -646.78 & 48 & 1.43s \\
816 & 139 & -646.43 & 46 & 3.31s & 11 & -646.81 & 48 & 1.56s \\
1 632 & 112 & -646.45 & 48 & 3.45s & 11 & -646.83 & 49 & 1.75s \\ 
3 264 & 99 & -646.59 & 48 & 3.83s & 7 & -646.77 & 49 & 2.19s \\
6 528 & 97 & -646.48 & 50 & 5.50s & 6 & -646.77 & 49 & 3.33s \\
13 056 & 93 & -646.37 & 48 & 7.99s & 7 & -646.84 & 49 & 5.99s \\
26 112 & 85 & -646.52 & 48 & 13.97s & 6 & -646.84 & 48 & 10.80s \\
52 224 & 98 & -646.49 & 50 & 25.39s & 5 & -646.83 & 49 & 25.01s \\
104 448 & 99 & -646.57 & 48 & 51.78s & 6 & -646.83 & 48 & 50.02s \\
\hline
\end{tabular}
\caption{For dataset A, the comparison of number of cycles required for convergence, the value of the objective function obtained, and time taken for various sizes of the initial parameter search space. Abbreviations: $LL$, the twice negative logarithm of the likelihood, also known as the objective function value.}
\label{tab:npod_npag}
\end{table}

\begin{figure}[H]
    \centering
    \includegraphics[width=0.63\columnwidth]{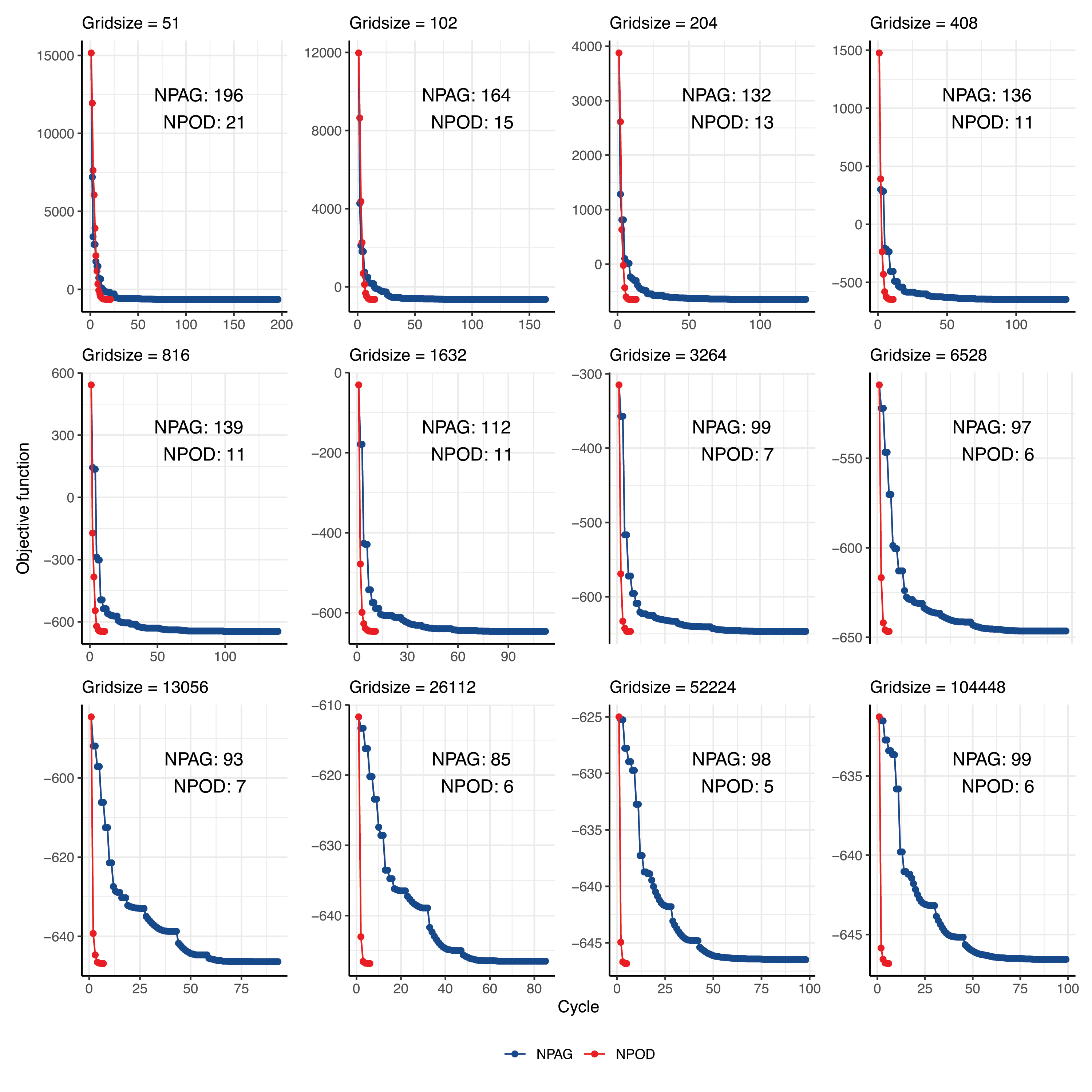}
    \caption{Comparison of objective function between NPAG and NPOD for different number of initial grid points in the parameter search space. Gridsizes are chosen as multiples of the number of subjects (n = 51).}
    \label{fig:objf_cycle_comparison_bimodal_ke}
\end{figure}

\begin{figure}[H]
    \centering
    \includegraphics[width=0.5\linewidth]{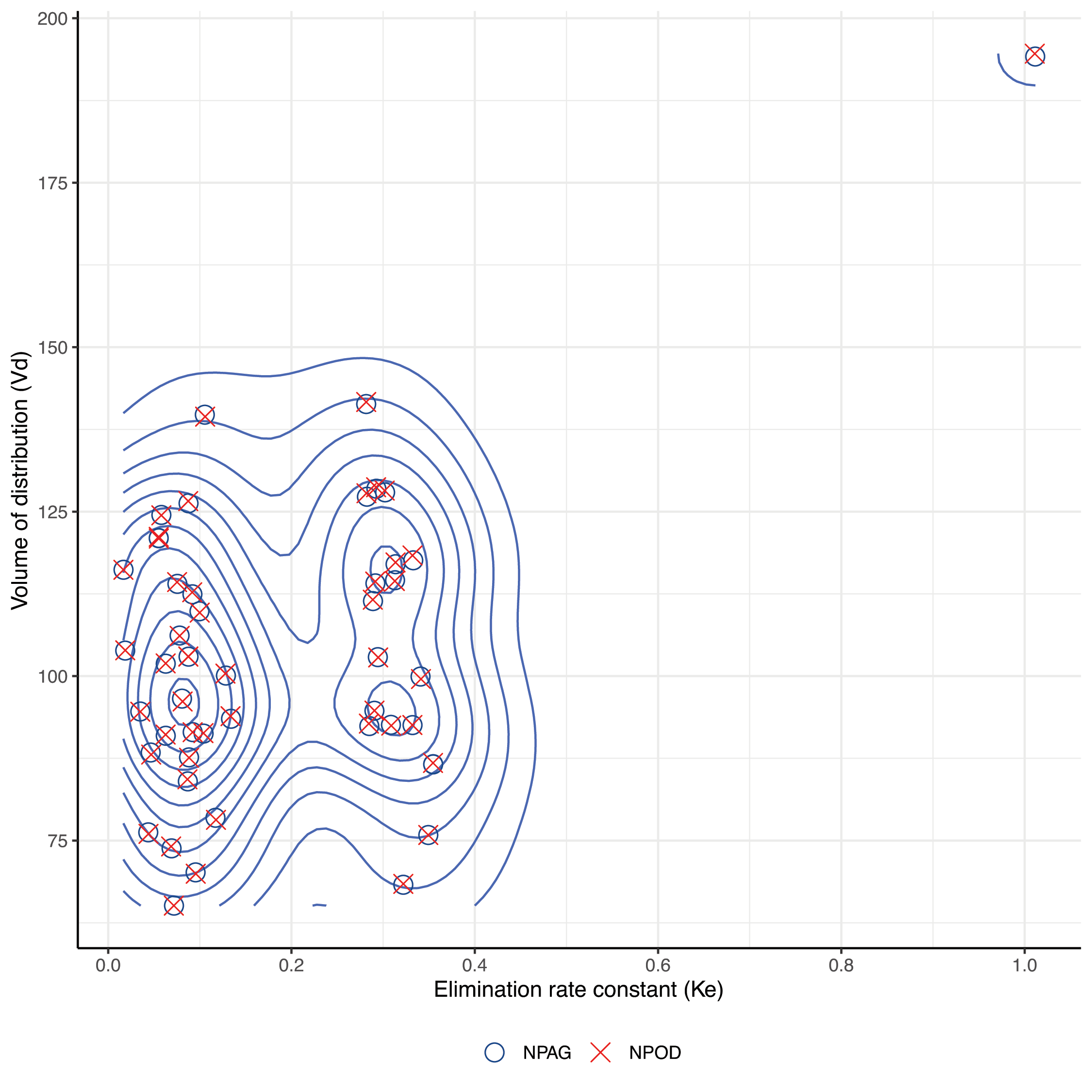}
    \caption{Kernel density estimate for the joint parameter distribution. The support points estimated by NPAG are shown as blue circles, and those by NPOD is shown as red crosses. The bimodal distribution of $K_e$ is readily apparent, with the univariate distribution of $V_d$, and the inclusion of an extreme outlier.}
    \label{fig:theta}
\end{figure}

Furthermore, the shape of the objective function across cycles is markedly different between NPOD and NPAG, shown in Figure \ref{fig:objf_cycle_comparison_bimodal_ke}. It is immediately apparent that NPOD has a much steeper convergence.

For dataset B, which to reiterate consists of real-world data, NPOD was able to obtain a solution as likely or more compared with NPAG with a lower number of cycles. However, for this dataset, the overall computation time was lower, with up to 5-fold difference, as seen in Table \ref{tab:table_two_eq_lag}.

\begin{table}[H]
\centering
\begin{tabular}{r|cccc|cccc}
\hline
& \multicolumn{4}{c|}{NPAG} & \multicolumn{4}{c}{NPOD} \\
\cline{2-9}
\textnumero & Cycles & $LL$ & Support points & Time & Cycles & $LL$ & Support points & Time \\
\hline
51 & 1091 & -337.93 & 19 & 68.43s & 189 & -331.98 & 19 & 47.74s \\
102 & 2166 & -337.97 & 20 & 95.11s & 248 & -336.91 & 17 & 36.87s \\
204 & 1234 & -336.56 & 20 & 75.95s & 119 & -342.64 & 17 & 32.75s \\
408 & 1313 & -336.54 & 19 & 71.25s & 68 & -346.98 & 18 & 21.31s \\
816 & 2218 & -343.91 & 20 & 116.91s & 74 & -345.31 & 17 & 21.72s \\
1 632 & 3034 & -343.91 & 20 & 110.95s & 69 & -345.38 & 17 & 22.62s \\
3 264 & 1415 & -343.92 & 20 & 65.83s & 192 & -335.75 & 18 & 36.29s \\
6 528 & 1913 & -343.84 & 20 & 73.75s & 75 & -336.66 & 18 & 25.46s \\
13 056 & 1401 & -337.91 & 20 & 77.60s & 85 & -336.09 & 17 & 36.40s \\
26 112 & 2169 & -336.54 & 20 & 122.14s & 82 & -334.35 & 17 & 56.66s \\
52 224 & 1209 & -336.53 & 20 & 135.22s & 65 & -335.45 & 18 & 89.48s \\
104 448 & 2014 & -336.53 & 20 & 202.92s & 75 & -335.38 & 18 & 147.74s \\
\hline
\end{tabular}
\caption{For dataset B, the comparison of number of cycles required for convergence, the value of the objective function obtained, and time taken for various sizes of the initial parameter search space. Abbreviations: $LL$, the twice negative logarithm of the likelihood, also known as the objective function value.}
\label{tab:table_two_eq_lag}
\end{table}

\begin{figure}[H]
    \centering
    \includegraphics[width=0.7\columnwidth]{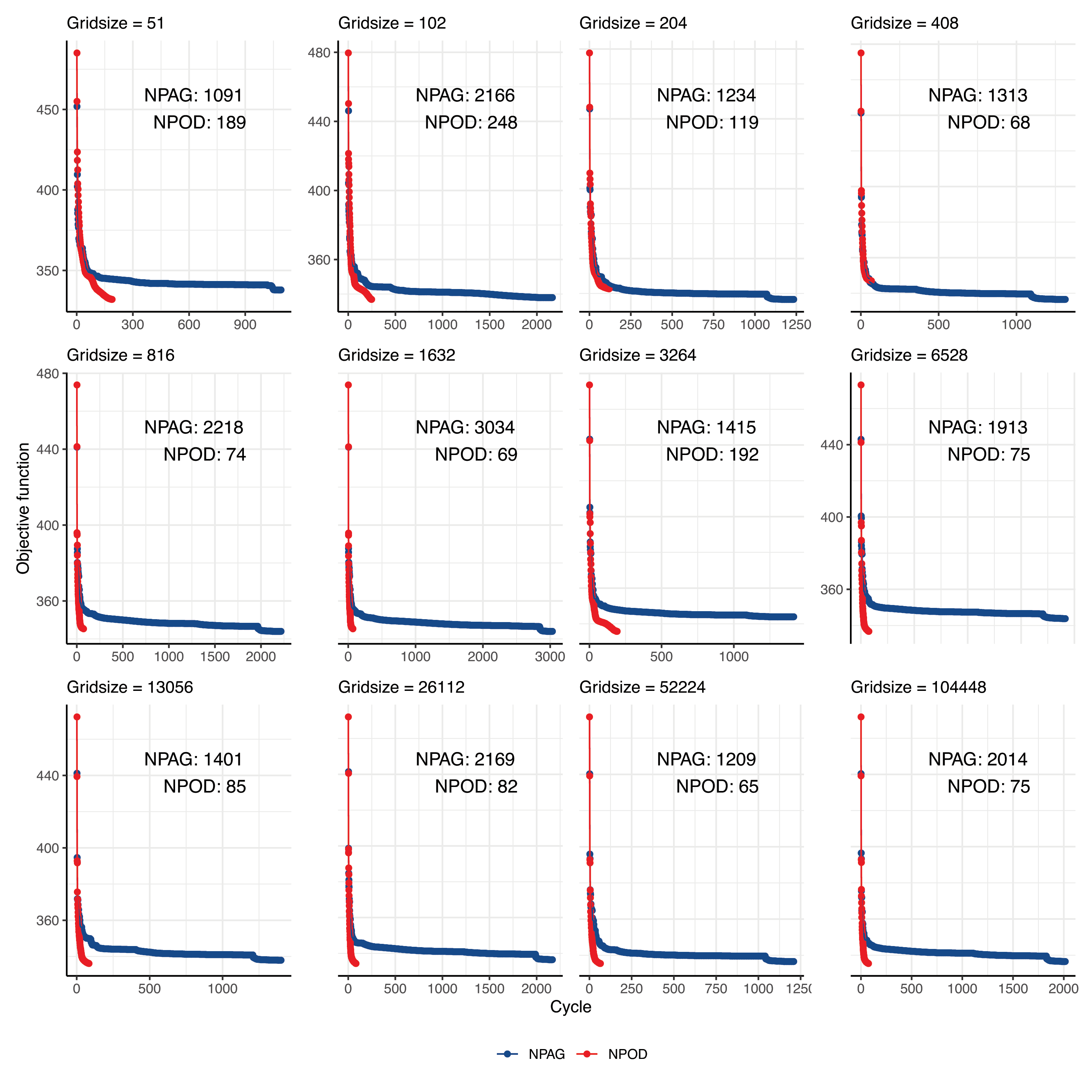}
    \caption{For dataset B, the comparison of objective function between NPAG and NPOD for different number of initial grid points in the parameter search space. For simplicity, the same grid sizes in the first example was used.}
    \label{fig:objf_cycle_comparison_two_eq_lag}
\end{figure}

The number of cycles required for convergence is again visualized for both algorithms in Figure \ref{fig:objf_cycle_comparison_two_eq_lag}.

\section{Discussion}
We have developed and demonstrated an algorithm for non-parametric parameter estimation with application to population pharmacokinetics. The algorithm relies on directional derivatives, which constitutes a new approach to parameter estimation in pharmacometrics.

Furthermore, we compared the NPOD algorithm to the current gold-standard non-parametric algorithm, NPAG on two datasets; one simulated without any noise in the observations, and another using real-world data. The two algorithms have some important differences, which is elucidated by the results in Table 1. Most importantly, NPOD was able to determine a solution that was as likely as that of NPAG. The time savings with NPOD is due to the markedly reduced number of cycles, with a difference of up to 20-fold for a simple model on a simulated dataset (dataset A), and more than twice that for a more complex model with real-world data (dataset B). However, the optimization step of NPOD, which is guided by the D-function value, is more expensive than the adaptive grid in NPAG, and as such requires more time in each cycle. Because we expect NPOD to always converge in fewer cycles than NPAG, we also expect that time to convergence will be at worst similar, and at best shorter than NPAG, although this remains to be empirically demonstrated as we gain experience with NPOD.

The estimated joint parameter distribution from both NPOD and NPAG was very similar, as seen in Figure 2, where the support points are close to perfectly overlapping. The simulated dataset included an extreme outlier, whose parameter values deviated greatly from the population weighted mean ($K_e$, 1.0 vs 0.187; $V_d$ 200.0 vs 103.8). This is especially impressive considering that the single outlier constitutes only 1/51, or approximately 2\% of the dataset. This quality is one of the many strengths of NPAG, which is also found for NPOD. The detection of outliers is an important aspect in population pharmacokinetics, and one of the chief advantages of non-parametric approaches, compared to parametric \cite{Goutelle2022}.

We evaluated the performance of the two algorithms over various densities of the initial search space. While this density does not appear to significantly affect the final objective function value for this simulated dataset, it does affect the number of cycle required to reach convergence. The "throw and catch" nature of NPAG ignores the gradient around the current local solution, which NPOD is sensitive to. Importantly, this gradient includes observation noise. NPAG is relatively insensitive to local gradient perturbations resulting from observation errors as it merely compares two potential and spatially distinct solutions at each cycle. However, NPAG is {\em completely naive} of the intervening space. In either case, both algorithms were capable of obtaining the most likely solution even from a very sparse initial parameter space, equal to the number of subjects. 

\edit{For Dataset A NPOD results in a lower log-likelihood (LL), indicating potentially more accurate results. A natural questions arises `` why use NPAG at all?''. Some distinctions in how the two algorithms operate might explain their differences. NPAG cycles are generally faster because they add points in a more straightforward manner, without necessarily checking if those points have been added before or if they are crucial to improving the solution. This can lead to quicker iterations but may include less relevant points.
In contrast, NPOD evaluates the likelihood surface more thoroughly, proposing new points that are specifically aimed at maximizing the objective function or minimizing the negative likelihood. As a result, every point added by NPOD is highly relevant to refining the outcome.
We are planning to investigate the conditions under which one algorithm would consistently be preferable over the other in a future work.}

The following procedure is proposed in Yamada at el. \cite{Yamada2021} for evaluating the global optimality of the final NPAG distribution and estimating its proximity to the optimum using the directional derivative $D(\Theta, F)$ defined above in \ref{eq:arab4.3.2} solely during the last NPAG iteration. As shown in Lindsay \cite{lindsay:gml83} and mentioned above in Methods section if $F^* = F^{ML}$ i.e. NPAG converged to a global maximum of a likelihood function then $\max \limits_{\xi \in \Omega}D(\xi,F^*)=0$. We propose the same evaluation steps for the final NPOD distribution and recommend calculating $\max \limits_{\xi \in \Omega}D(\xi,F)$ only at the end of the algorithm using deterministic or stochastic optimization methods. 

\edit{\paragraph{Future work:}}
 
\edit{In our experience thus far, NPOD converges at a faster rate when compared with NPAG for a larger population with a given model and prior. In the era of increasingly larger datasets, NPOD may therefore prove advantageous. However, it is not a given that NPOD will always outperform NPAG; therefore future work will include additional comparisons between NPAG and NPOD to contrast and clarify specific scenarios when one algorithm should be preferred.}

\edit{We have also observed that the performance of the NPOD algorithm depends on the initial grid or prior. The closer it is to the true solution the faster the convergence of NPOD. We are planing to explore the ways to improve initial grid point in the future.}

\edit{At present, formal analyses of convergence guarantees and algorithmic complexity for both NPAG and NPOD are limited. NPOD, being a gradient-based approach, benefits from some theoretical convergence guarantees, particularly when the likelihood surface is smooth and wellbehaved. Under these conditions, it can converge more efficiently to the maximum likelihood solution when compared with the adaptive grid in NPAG. However, the complexity of the models and datasets, such as in the case of noisy or high-dimensional data, introduces challenges in predicting convergence behavior. For NPAG, the lack of formal convergence guarantees is more pronounced, as its heuristic nature can result in variability in performance depending on the problem being solved. This makes it difficult to derive generalizable results regarding its runtime or convergence. While NPOD has shown promising improvements in speed and accuracy, a deeper theoretical analysis, especially with respect to different types of datasets and models, is something we aim to explore in future studies.}

\vspace{10mm}

\section{Conclusion}
We have developed and demonstrated a new algorithm, NPOD, for non-parametric parameter estimation with application to population pharmacokinetics. The algorithm was able to estimate the population joint parameter distribution as accurate as NPAG, but requires far fewer cycles to reach convergence. An application of directional derivates represents an important step forward in both the development and application of non-parametric approaches in pharmacometrics.


\section{Competing Interests}
Michael N. Neely, Julian Otalvaro and Walter M. Yamada were partially supported by NIH-NIAID R01AI173238. The authors have no conflicts of interest to declare that are relevant to the content of this chapter.

\bibliographystyle{plain}
\bibliography{Kryshchenko_main}

\newpage

\section{Appendix}
\begin{algorithm}[h]
\setlength\baselineskip{14pt}
\caption{Non-parametric Optimal Design (NPOD) algorithm. \\
Input: $( \bm{Y}, \bm{\phi}^0, \bm{a}, \bm{b},\bm{t},\Delta_D,\Delta_F, \Delta_e,\Delta_\lambda)$,
$\bm{a}$ and $\bm{b}$ are the lists of lower and upper bounds, respectively, of $\Theta$, $\bm{t}$ is the number of Nelder-Mead iterations;
$\Delta_D$ is the minimum distance allowable between points in the estimated $F^{ML}$.
Output: $(\bm{\phi}, \bm{\lambda}, l(\bm{\lambda}, \bm{\phi})  )$.
}
\label{OuterLoopOfNPAG}
\begin{algorithmic}[1]
\Procedure{NPOD}{$\bm{Y}$, $\bm{\phi}^0, \bm{a}, \bm{b},\Delta_D$} \Comment{Estimate $F^{ML}$ given $\bm{Y}$}

 \noindent \State {Initialization:
	$\bm{\phi} =\bm{\phi}^0$,
	$LogLike=-10^{30}$, $\Delta_F=10^{-2}$, $\Delta_L=10^{-4}$, $\Delta_\lambda=10^{-3}$,
	$n = 0$}
	\While{True}

		\State{Calculate $\bm{\Psi}(\bm{\phi})$}
			\Comment{$N \times K$ matrix $\{ p(Y_i\vert\phi_k)\}$}
	
		\State {$[\bm{\hat{\lambda}}(\bm{\phi}),l(\bm{\hat{\lambda}}(\bm{\phi})] \longleftarrow \mbox{PDIP}(\bm{\Psi}(\bm{\phi})) $ }
			\Comment{for PDIP see \cite{Yamada2021}} 
		\State {$\bm{\phi} \longleftarrow \mbox{CONDENSE}(\bm{\phi},\bm{\hat{\lambda}}(\bm{\phi}), \Delta_\lambda)$}
			\Comment{Alg. 2}

        \State{[$\bm{\phi}, \bm{\Psi}(\bm{\phi})] \longleftarrow \mbox{REDUCE}(\bm{\Psi}(\bm{\phi}), \bm{\phi}) $} \Comment{Alg. 3}

		\State {$[\bm{\hat{\lambda}}(\bm{\phi}),l(\bm{\hat{\lambda}}(\bm{\phi})] \longleftarrow \mbox{PDIP}(\bm{\Psi}(\bm{\phi})) $}
			\Comment{$\mbox{PDIP returns } G^{n}$ ( \cite{Yamada2021})}

		\State {$NewLogLike =  l(\bm{\hat{\lambda}}(\bm{\phi}),\bm{\phi})$}

        \If{ $ \mid LogLike - NewLogLike \mid > \Delta_F $}
			\State{\Return $[ \bm{\phi}, \bm{\lambda}, NewLogLike]$}
		\EndIf
     
		\If{ $ \left( n > \mbox{MAXCYCLES} \right) $}	
			\State{\Return $[ \bm{\phi}, \bm{\lambda}, NewLogLike]$}
		\EndIf

  \State{$\bm{\phi} \longleftarrow \mbox{Dopt}(\bm{\phi},\bm{\lambda},\bm{\Psi}(\bm{\phi}),\bm{a},\bm{b},\bm{t}, \bm{\Delta_D}) $} \Comment{Alg. 4}
  \State{$n \longleftarrow n + 1$}
		\State{$LogLike \leftarrow NewLogLike$ }		
\EndWhile

\EndProcedure
\end{algorithmic}
\end{algorithm}

\begin{algorithm}
\setlength\baselineskip{21.5pt}
\caption{Condense algorithm.
	Input:  $(\bm{\phi},  \bm{\lambda}, \Delta_\lambda)$, Output: $\bm{\phi}^c$ \\
        Note: $\bm{\phi}^c$ is considered a subset of $ \bm{\phi}$   
        }\label{Algorithm2}

\begin{algorithmic}
\Function{Condense}{$\bm{\phi},\bm{\lambda}, \Delta_\lambda$}

\State{ {\bf ind} = {\bf find} ( $\bm{\lambda} > (\max\bm{\lambda})\Delta_\lambda $ )} \Comment{ Inequality and max are performed component-wise}

\State{ $\bm{\phi}^c = \bm{\phi}(: , {\bf ind})$}

\State { return $\bm{\phi}^c$ }

\EndFunction
\end{algorithmic}
\end{algorithm}

\begin{algorithm}
o\setlength\baselineskip{21.5pt}
\caption{Reduce Algorithm.
	Input:  $(\bm{\Psi(\phi)}, \bm{\phi})$, Output: $\bm{\phi}, \bm{\Psi}(\bm{\phi})$ 
 Note: both $\bm{\Psi(\phi)}$ and $\bm{\phi}$ are subsets of the ones used as input.   
}\label{Algorithm3}

\begin{algorithmic}
\Function{Reduce}{$(\bm{\Psi(\phi)}, \bm{\phi})$}	

\State{$n\psi(\phi) = norm(\Psi(\phi)) $}
\State{ $(r,perm) = QR(n\psi(\phi))$ }
\State{ $keep = []$ }
\State{ \textbf{for} $i .. ncol$ }
\State{ \ \ \ $ratio = r[i,i]/norm(r[:,i])$ }
\State{ \ \ \ \textbf{if} $|ratio| > 1e-8$ \textbf{push}$(perm[i])$ to \textbf{keep} }
\State{ \textbf{end for} }

\State{ $\phi = \phi[keep,:]$ }
\State{ $\psi = \psi[:,keep]$ }
\State{ \textbf{return} $(\psi, \phi)$ }

\EndFunction
\end{algorithmic}
\end{algorithm}

\begin{algorithm}
\setlength\baselineskip{21.5pt}
\caption{Dopt algorithm
	Input:  $(\bm{\phi},\bm{\lambda},\bm{\Psi}(\bm{\phi}),\bm{a},\bm{b},\bm{t},\bm{\Delta}_D)$, Output: $\bm{\phi}$ 
}\label{Algorithm4}

\begin{algorithmic}
\Function{Dopt}{$\bm{\phi},\bm{\lambda},\bm{\Psi}(\bm{\phi}),\bm{a},\bm{b},\bm{t}$}	
			
\For{$k = 1:K$} 

\State{ $\bm{\phi}_k = argmax^{t}_{\bm{\xi} \in \bm{\Omega}}(D(\bm{\xi},\bm{\lambda},\bm{\Psi}))$} \Comment{ see formula for $D$ in Eq. (\ref{nes_sup})}

\For{$ink = 1:\text{dimension}(\bm{\phi}_k)$}
    \State $\text{new\_dist} = \sum \frac{|\bm{\phi}_k - \bm{\phi}(:,ink)|}{b-a}$
    \State $\text{dist} = \min(\text{dist}, \text{new\_dist})$
\EndFor
\State $\text{up} = \text{sign}(\min(\bm{\phi}_k - a'))$
\State $\text{down} = \text{sign}(\min(b' - \bm{\phi}_k))$
\If{$(\text{dist} > \bm{\Delta}_D) \wedge (\text{up} > -1) \wedge (\text{down} > -1)$}
    \State $\bm{\phi} = [\bm{\phi}, \bm{\phi}_k]$
\EndIf
\EndFor
\State { return $\bm{\phi}$ } 
 
\EndFunction
\end{algorithmic}
\end{algorithm}

\vspace{10pt}

\newpage

\end{document}